\documentclass[a4paper,twocolumn]{esapub}
\usepackage{ifluatex,ifxetex} 
\ifluatex
\usepackage{fontspec}
\else\ifxetex
\usepackage{fontspec}
\else
\usepackage{times}
\fi

\usepackage[numbers]{natbib}
\usepackage{graphicx}
\usepackage{comment}
\usepackage{hyperref}

\title{Using all-sky optical observations for automated orbit determination and prediction for satellites in Low Earth Orbit}
\author[1,2]{T.P.G. Wijnen}
\author{R. Stuik}
\author{M. Rodenhuis}
\author{M. Langbroek}
\affil{Leiden Observatory, Leiden University, PO Box 9513, 2300 RA Leiden, The Netherlands,\newline Email: wijnen@strw.leidenuniv.nl}
\author{P. Wijnja}
\affil{Space Security Center, Royal Netherlands Air Force, Breda, The Netherlands}

\begin{document}


\maketitle

\begin{abstract}
We have used an existing, robotic, multi-lens, all-sky camera system, coupled to a dedicated data reduction pipeline, to automatically determine orbital parameters of satellites in Low Earth Orbit (LEO).  Each of the fixed cameras has a Field of View of 53 x 74 degrees, while the five cameras combined cover the entire sky down to 20 degrees from the horizon. Each of the cameras takes an image every 6.4 seconds, after which the images are automatically processed and stored. We have developed an automated data reduction pipeline that recognizes satellite tracks, to pixel level accuracy ($\sim$ 0.02 degrees), and uses their endpoints to determine the orbital elements in the form of standardized Two Line Elements (TLEs). The routines, that use existing algorithms such as the Hough transform and the Ransac method, can be used on any optical dataset. 

For a satellite with an unknown TLE, we need at least two overflights to accurately predict the next one. Known TLEs can be refined with every pass to improve collision detections or orbital decay predictions, for example. For our current data analysis we have been focusing on satellites in LEO, where we are able to recover between 50\% and 80\% of the known overpasses during twilight. We have been able to detect LEO satellites down to 7$^{\rm{th}}$ visual magnitude. Higher objects, up to geosynchronous orbit, were visually observed, but are currently not being automatically picked up by our reduction pipeline. We expect that with further improvements to our data reduction, and potentially with longer integration times and/or different optics, the instrumental set-up can be used for tracking a significant fraction of satellites up to geosynchronous orbit.  
\end{abstract}

\section{Introduction}

The continuous monitoring and tracking of objects in Earth's vicinity, both Earth-orbiting and Near Earth Objects (NEOs), is crucial to assess potential threats to assets in space and life on Earth. Historical events have demonstrated that the consequences of a NEO entering our atmosphere and impacting on Earth can be devastating \citep{krinov63}. Although the probabilities of such events are small, they do pose a long-term statistical hazard \citep{chapman94}. Simultaneously, our technology-driven society crucially depends upon space infrastructure for, among others, communication, navigation, (bank) data transfer and earth monitoring. Loss, even partially, of these space-dependent services could lead to chaotic and possibly life-disrupting situations \citep{office13}.
As the amount of human-made debris orbiting Earth grows, the possibility of a collision with (operational) spacecraft or other objects increases \citep{liou08}. A collision would generate more debris, that may in turn cause a chain-reaction of collisions, known as the Kessler syndrome \citep{kessler78}, which would endanger both our assets in and access to Low Earth Orbit (LEO). 

It is therefore of vital importance to continuously survey the sky, detect, catalogue and track Earth satellites and NEOs, and determine their orbits. The orbital parameters have to be updated frequently to make sure that orbital calculations are accurate and reliable, and that resulting collision avoidance manoeuvres and impact hazard assessments are adequate. Ideally, these monitoring processes are continuous, automated and performed by sensors with all-sky coverage. 

Here we present a data reduction pipeline that can automatically detect tracks of LEO satellites in optical data, and determine the orbital parameters of those satellites. With every observed successive passage of a satellite, its orbital elements, generally published as Two Line Elements (TLEs), can be refined. The pipeline was written with data from an existing instrument that covers almost the entire sky and has been developed with off-the-shelf components. 

\begin{figure*}[!ht]
  \centering
  \begin{tabular}{ccc}
  \includegraphics[width=.5\linewidth]{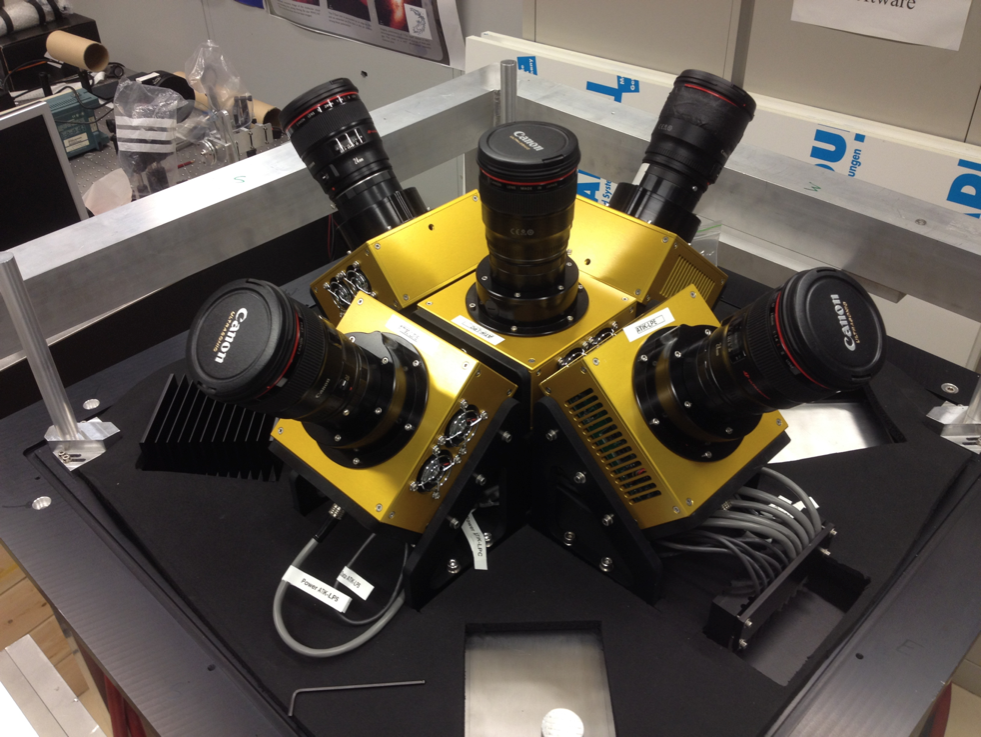} &
  \includegraphics[width=.5\linewidth]{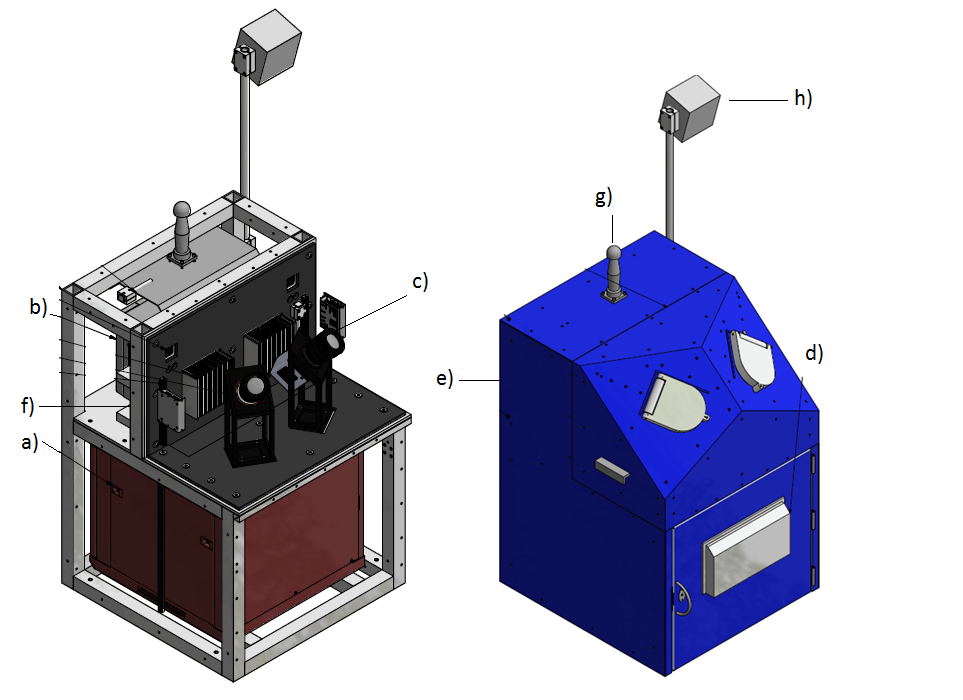}
  \end{tabular}
  \caption{The northern hemisphere MASCARA instrument (left, image taken from \citep{talens17}) with five ATIK 11000M sensors and 24 mm $f/1.4$ Canon lenses, and a schematic of the bRing enclosure (right, image taken from \citep{stuik17}), with and without cover panels. bRing is equipped with the same lenses and two FLI Microline ML11002M sensors.}
  \label{fig:instruments}
\end{figure*}

\section{Instrument}

The Multi-site All-Sky CameRA (MASCARA, see Figure \ref{fig:instruments}) is a robotic instrument with the main objective to detect transiting exoplanets through photometric observations of stars with visual magnitude $4 < m_{V} <  8$ \citep{talens17}. The system consists of five fixed cameras, each having a Field of View of $53 \times 74$ degrees, and a $24 \times 36$ mm interline Charged Coupled Device with 9 $\mu$m pixels. Together the cameras provide coverage of the entire sky down to 20 degrees from the horizon, and down to $m_{V} = 8.3$ (and potentially to $m_{V} = 10$). There are two MASCARA stations to cover both the northern and southern hemisphere, located in La Palma (Canary Islands, Spain) and ESO's La Silla Observatory (Chile) respectively. Images with an exposure time of 6.4 seconds are taken at 6.4 seconds cadence, and are automatically processed and stored with a time stamp. The cadence and instrument are optimised for the main science case, while the observation of satellites is an inevitable by-product. Nonetheless, the results presented here demonstrate that a similar facility would be a promising candidate for dedicated monitoring of artificial bodies in (low) Earth orbit.

The images we used for the development of our data reduction pipeline were taken by the bRing instrument, which is essentially a smaller version of MASCARA with two fixed cameras (see Figure \ref{fig:instruments}) and an observation strategy that includes taking alternating long (6.4 seconds) and short (2.5 seconds) exposures at 12.8 seconds cadence. It was developed to monitor the transit of a hypothesised ring system around the planet $\beta$ Pictoris b \citep{stuik17}. We used the data from the station in Sutherland, South Africa. MASCARA and bRing use the same software suite; our routines can therefore directly run on data from both instruments, and in principle on any optical dataset. 

\section{Data reduction}

\begin{figure*}
  \centering
  \begin{tabular}{ccc}
  \includegraphics[width=.5\linewidth]{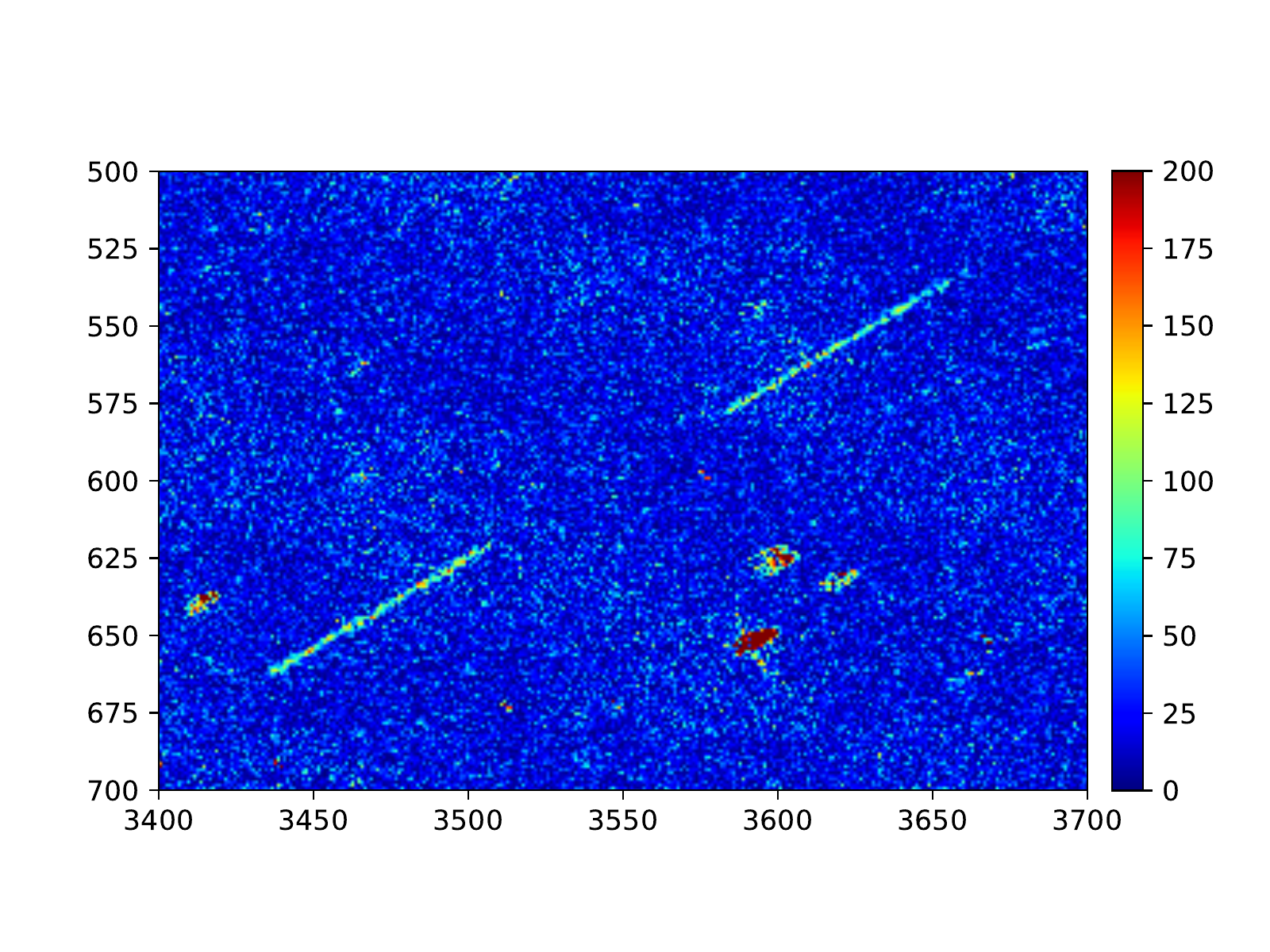} &
  \includegraphics[width=.5\linewidth]{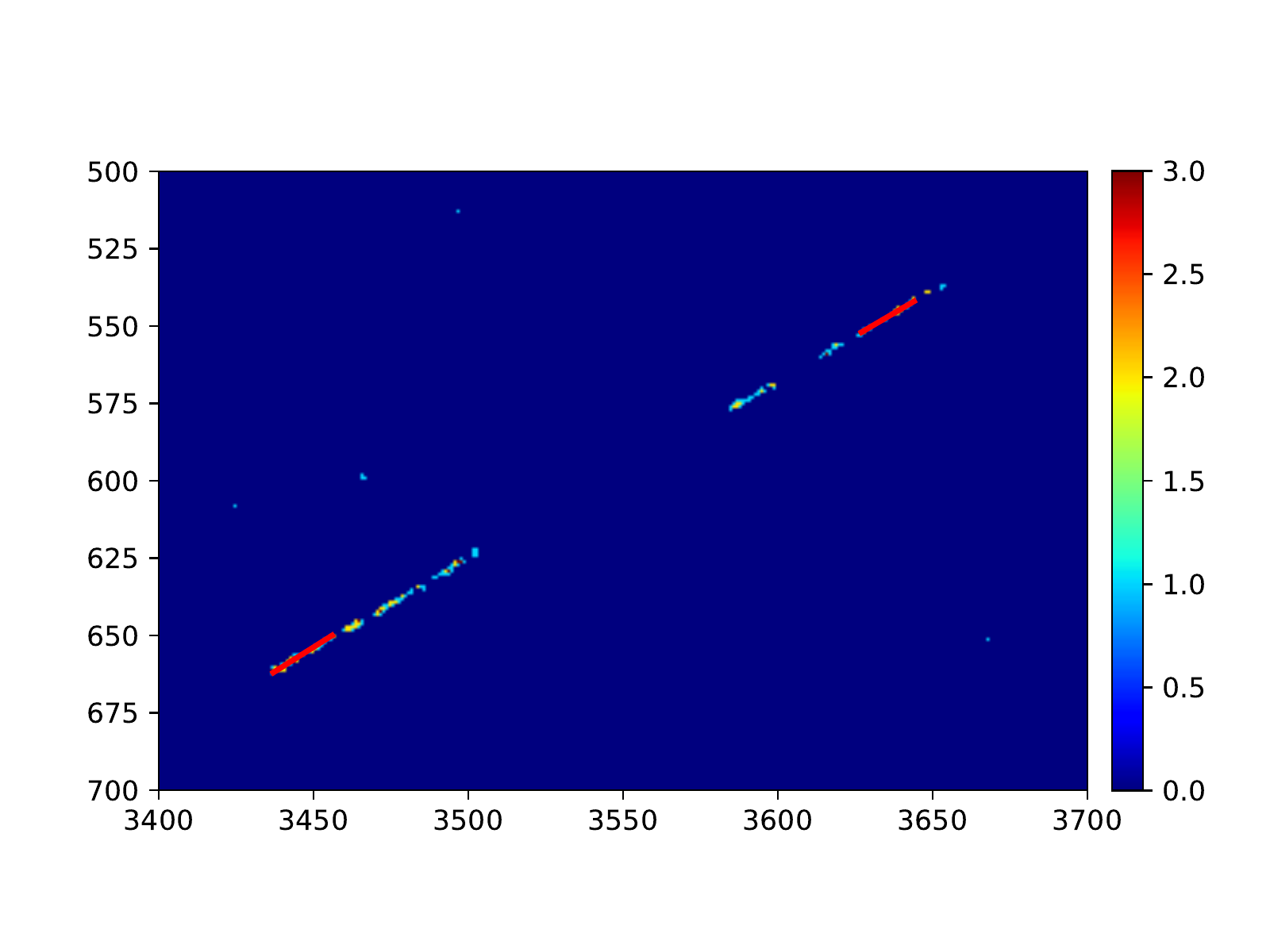}
  \end{tabular}
  \caption{\emph{Left:} the result of aligning two consecutive images, subtracting them and taking the absolute value of the difference. \emph{Right:} outcome of our enhance procedure in which we make satellite tracks stand-out from the background. The gaps result from the satellite's movement in front of several stars. The red lines are line segment recognised by the Hough transform. The colour scale indicates the pixel values, while the x- and y-axis give the pixel number.}
  \label{fig:imagereduction}
\end{figure*}

We use the already existing astrometic procedures from the MASCARA data pipeline \citep{talens18} to align, and subtract two consecutive images. The left image of Figure \ref{fig:imagereduction} shows two consecutive tracks of a satellite, resulting from this subtraction. Stars (with small variations in detected brightness due to inter-pixel sensitivity variations) are still visible in this image and we mask them with the median pixel value of the whole image, using the same astrometric routine. Subsequently we enhance the contrast between satellite tracks and the background noise with various operations in which we focus on pixels with a value that is 2.5 standard deviations above the mean pixel value, $\mu_{\rm p}$, and have at least six neighbouring pixels that are 1 standard deviation above $\mu_{\rm p}$. The end result is shown in the right image of Figure \ref{fig:imagereduction}. The satellite track crosses over several stellar residuals, with the masking process resulting in some gaps in the track

\subsection{Detection of tracks}

The automated recognition of line segments, i.e. satellite tracks, is done with the Hough transform \citep{hough62, duda72} of the OpenCV Python package \citep{bradski????} on the data points that remain in the reduced image. The Hough transform is able to detect multiple line segments in a single image. In our current implementation, line segments must have a minimal length of 15 pixels and a gap in a continuous line cannot be larger than 3 pixels. These parameters focus the selection routine on satellites in LEO and prevent the detection of objects in higher orbits, e.g. geosynchronous and Medium Earth Orbit. The right panel of Figure \ref{fig:imagereduction} shows that an intermittent satellite track is not always recognized as a single track by this routine. We therefore only use the Hough transform to locate all the satellite tracks in a reduced image. The determination of their orientation, begin- and end-point is done with the RANSAC method \citep{fischler81}. Contrary to the Hough transform, the RANSAC method can only fit one line through a set of data-points and cannot handle multiple tracks in a single image.

\begin{figure}
\centering
\includegraphics[width=\linewidth]{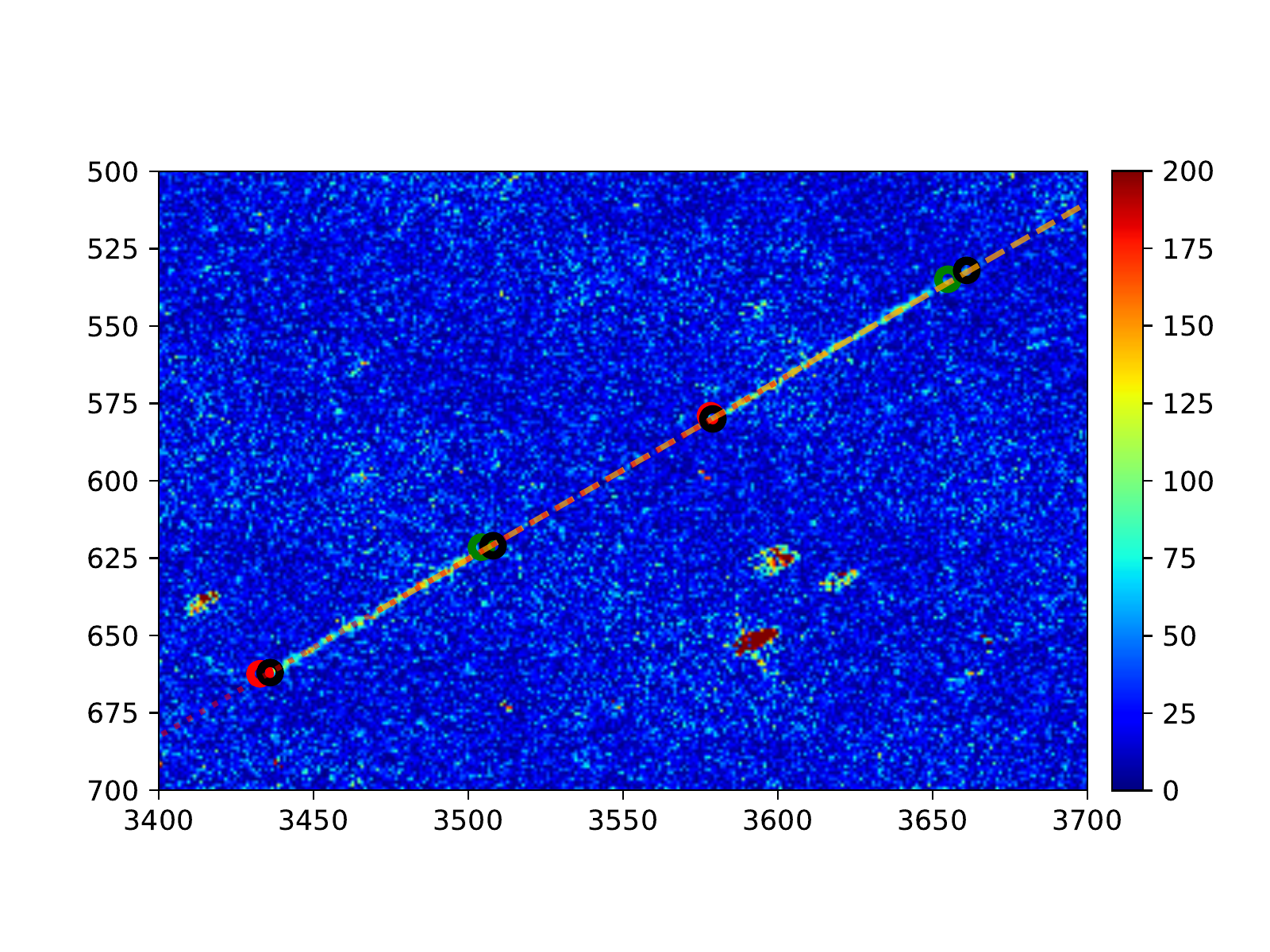}
\caption{Identical to Figure \ref{fig:imagereduction} but with the RANSAC line fits plotted as dashed and dotted lines and the determined endpoints as black open circles. The green and red open circles give the start and endpoints respectively as calculated from the TLE of this satellite provided by the CSpoC database \citep{????-1}. }\label{fig:image_endpoints}
\end{figure}

\begin{figure*}
  \centering
  \begin{tabular}{ccc}
  \includegraphics[width=.5\linewidth]{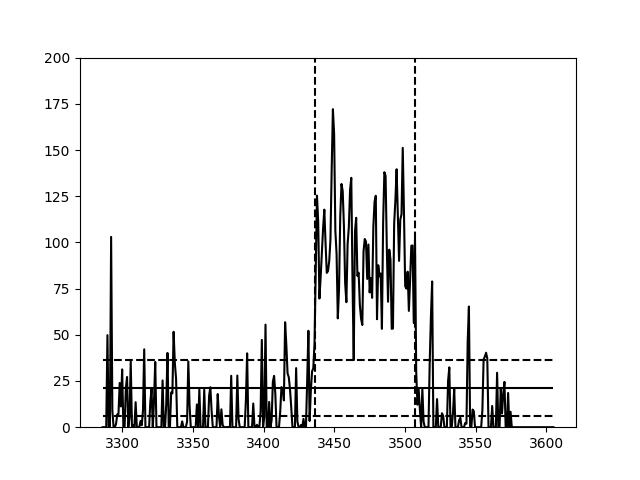} &
  \includegraphics[width=.5\linewidth]{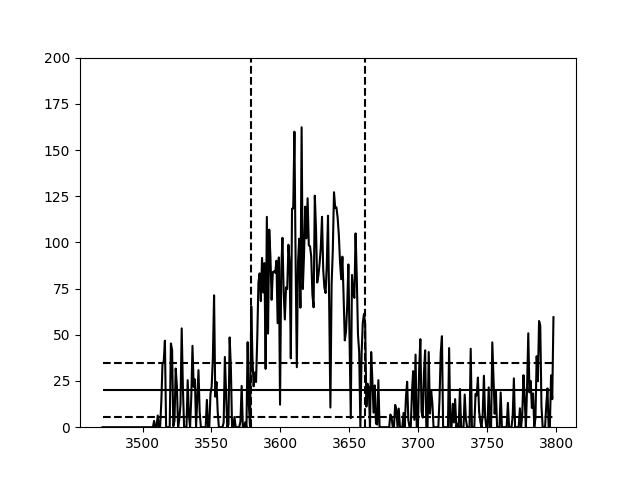}
  \end{tabular}
  \caption{The absolute pixel values along the line fits from the RANSAC method for the tracks in Figure \ref{fig:image_endpoints}. The x-axis corresponds to the x-position in Figure \ref{fig:image_endpoints}. The horizontal solid line gives the mean pixel value, while the horizontal dashed line are plus and minus 1 standard deviation. The vertical dashed lines indicate the endpoints found by the merit functions.}
  \label{fig:endpoint_determination}
\end{figure*}

We select an area with a pre-defined size around each found line segment in the reduced image and apply the RANSAC method to the data points within this area. The RANSAC method is capable of fitting a line through a whole track even if it has gaps. The dotted and dashed line in Figure \ref{fig:image_endpoints} shows the lines that are found in this way for the tracks in Figure \ref{fig:imagereduction}. By overlaying these fitted lines on the subtracted image, we can determine the begin- and end-point from the (absolute) pixel values along the satellite track. We start at a pixel that was also part of the line segment found by the Hough transform and has the highest pixel value. We define the following merit function: going along the fitted line in each direction from the starting pixel, we cumulatively add the difference between each pixel value and the mean plus standard deviation of the local pixel values. We define the end points as the maximum of each merit function. Figure \ref{fig:endpoint_determination} shows the pixel values and the endpoints determined in this way for the satellite tracks in Figure \ref{fig:imagereduction}. In Figure \ref{fig:image_endpoints}, these endpoints are plotted as black open circles and can be compared to the endpoints derived from the CSpoC database TLE \citep{????-1}. With these endpoints and the associated time stamps of the images, we can determine the orbital elements of the satellite.

\subsection{Determination of orbital elements}

With these determined (sky) positions at known times, we can derive the orbital elements of satellites with an unknown TLE. Applying the Gaussian method \citep{curtis14} to three consecutive sky positions with associated times, provides an initial estimate of the state vector for the middle sky position. The state vector is converted to the six classical orbital elements with Kepler's equations. With more than three sky positions, we can repeat this procedure and average the found orbital elements. This only gives a reliable estimate for the inclination, right ascension of the ascending node and the mean motion. 
With these initial estimates, we can refine the orbital elements using all found sky positions at once.

\subsection{Refinement of orbital elements}

We refine the TLEs for satellites in LEO with a publicly available SGP4 Python package \citep{rhodes????} and the parameters from \cite{vallado06}. Using a least squares method, we iteratively change the orbital elements to minimize the difference between the sky positions predicted by the SGP4 propagation and the sky positions that we have determined from the images. For satellites with an unknown TLE, we first only vary the eccentricity, argument of periapsis and mean anomaly, while keeping the other three elements fixed, for a better estimate of those parameters. Subsequently, we treat all six orbital elements as variable. For satellites with known TLEs, we first precess the right ascension of ascending node, argument of periapsis and mean anomaly to the epoch of observation, and then perform the least squares routine with all orbital elements as variables. For the development of the least squares routine, we did not use the sky positions as we have determined them, but as they are given by the CSpoC TLE. This allowed us to verify and optimize the least squares routine, before applying it to the possibly slightly offset sky positions that we determine. 

\begin{figure*}
  \centering
  \begin{tabular}{ccc}
  \includegraphics[width=.5\linewidth]{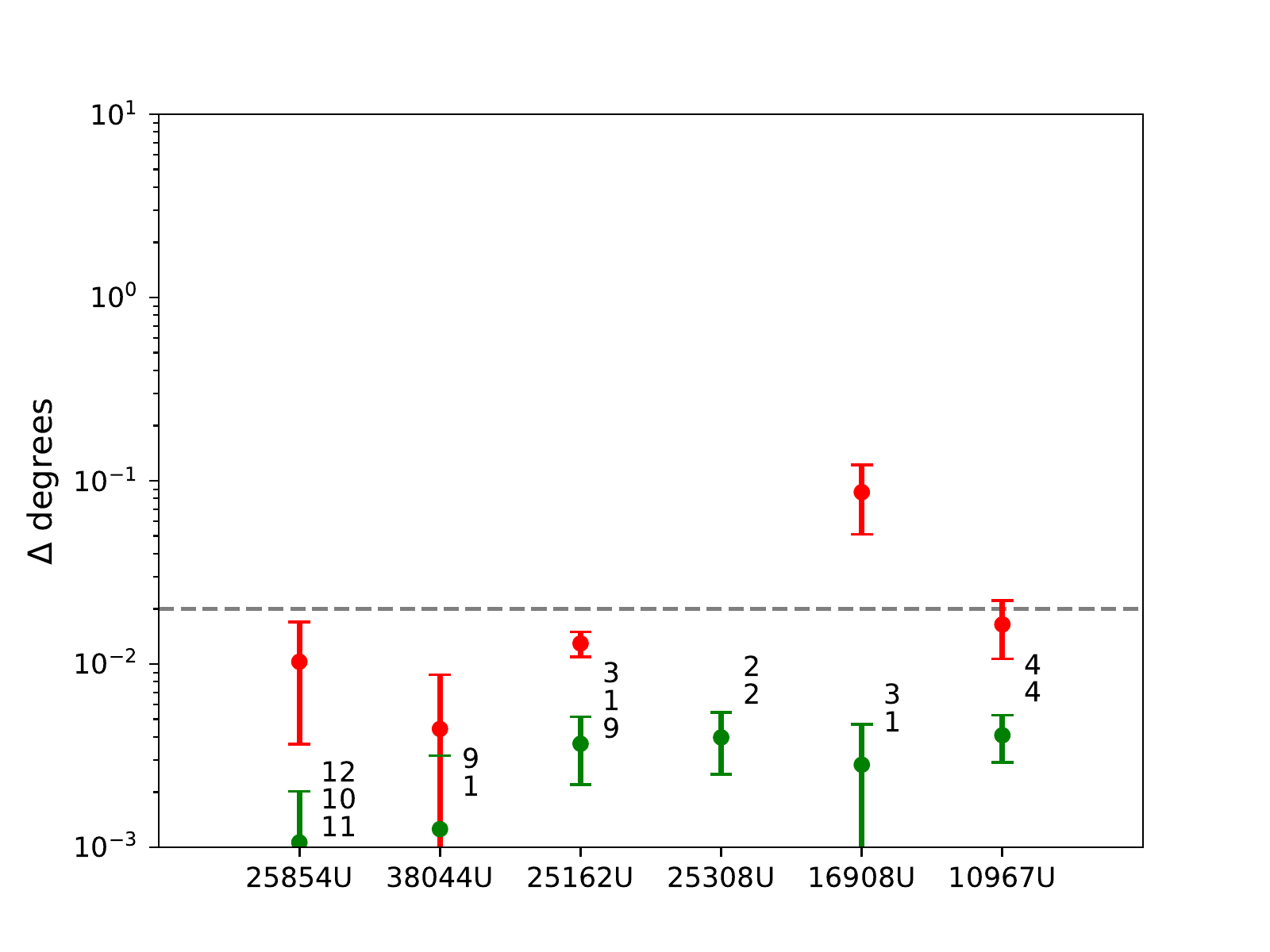} &
  \includegraphics[width=.5\linewidth]{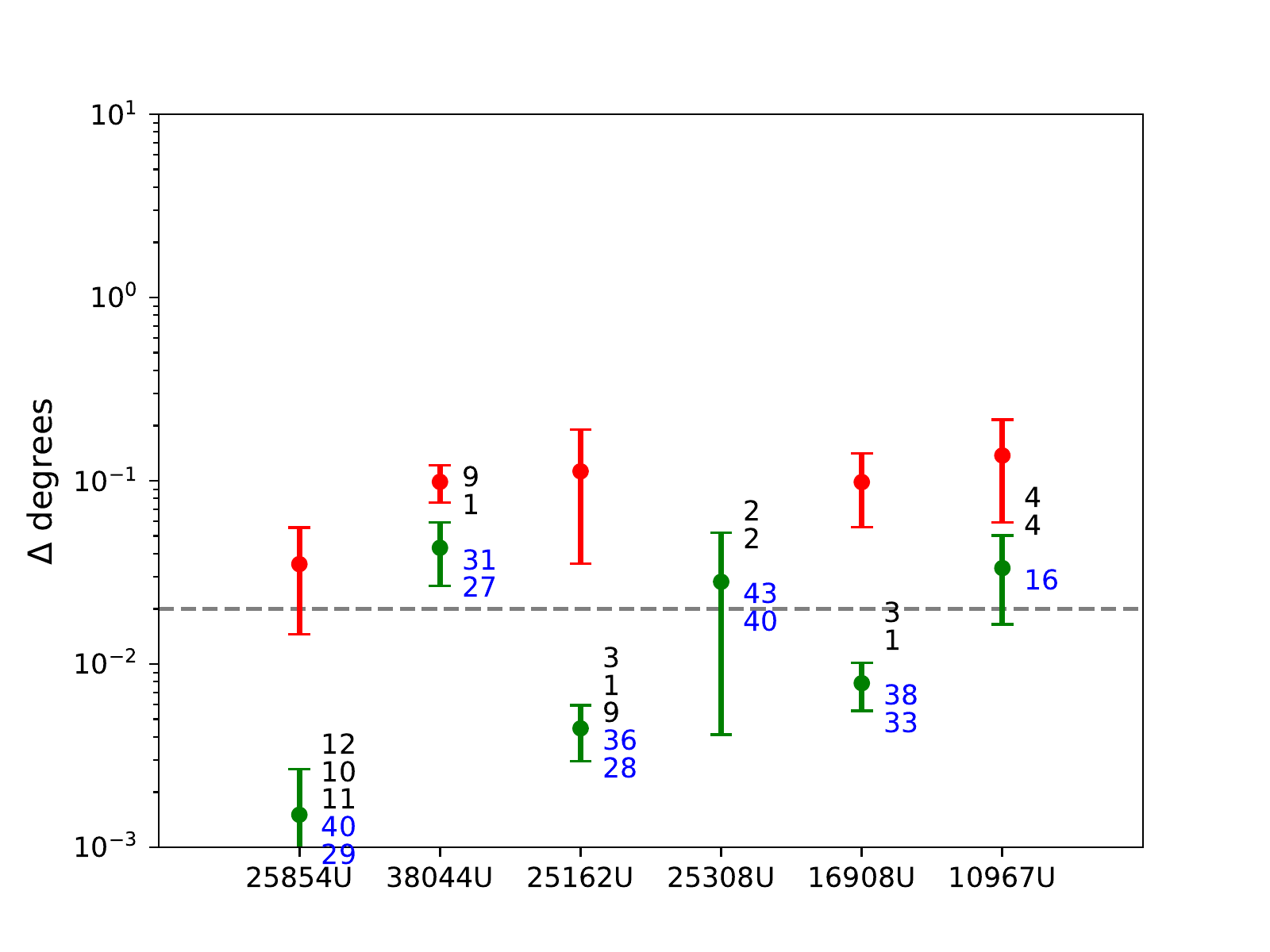}
  \end{tabular}
  \caption{The mean deviation on the sky in degrees between the `observed' positions and those predicted by our derived TLE for six satellites after one night \emph{(left)} and two consecutive nights \emph{(right)}. We used the positions expected from the CSpoC TLE, e.g. green and red open circles in Figure \ref{fig:image_endpoints}, as observed. This image therefore shows the performance of the least squares routine, rather than our complete pipeline. 
The green points correspond to the method in which we use the known TLEs, while the red points corresponds to the method in which we treat the TLE as unknown. The numbers are the sky positions used for each passage during the first (black) and second (blue) night. For example,  we only had two sky positions for each passage of satellite 25308U, and we were therefore not able to perform an initial orbital determination with the Gauss method. In this case, the mean deviation is averaged over the deviation of four sky positions. The horizontal dashed line corresponds to the size of a pixel on the sky, i.e. approximately the lower limit of the accuracy that we can obtain.}
  \label{fig:orbit_determination}
\end{figure*}

In figure \ref{fig:orbit_determination} we show the performance of our least squares routine for six satellites, using data from one night (left) and two consecutive nights (right). We used satellites for which we were able to determine their sky positions within 10 pixels from the position expected from the CSpOC TLE. For the first night, both the routine for known and unknown TLEs, indicate that the least square routine itself is accurate enough for our pixel resolution of 0.02 degrees. When the data of two consecutive nights is used, we see that the deviations are dominated by the large number of datapoints in the second night. Additionally, we did not take into account the elevation of the satellites above the horizon in the analysis. Possibly, some satellites passed at higher altitudes during the second night. 

We note that if only a single passage it used to predict the positions of the next overflight, the mean deviation for the second passage would be several degrees in case we assume that the TLE is unknown. If we assume that the TLE is known, the method performs as accurate as illustrated in Figure \ref{fig:orbit_determination}.

\subsection{Selectivity}

\begin{figure*}
  \centering
  \begin{tabular}{ccc}
  \includegraphics[width=.5\linewidth]{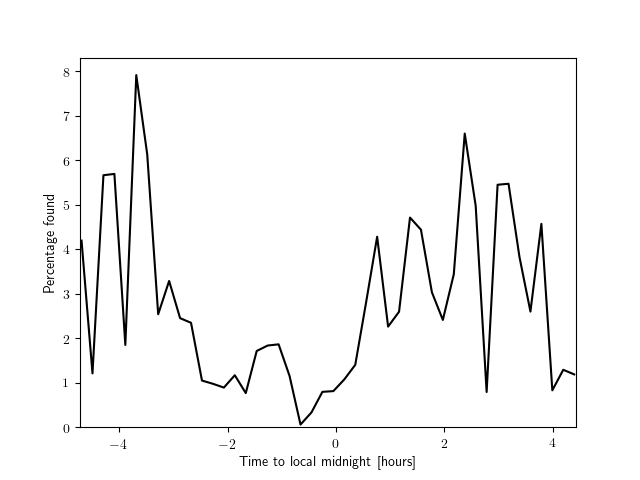} &
  \includegraphics[width=.5\linewidth]{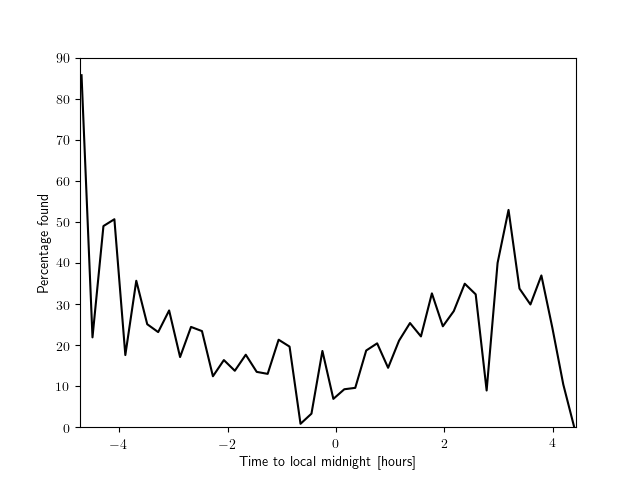}
  \end{tabular}
  \caption{The fraction of detected satellites in the field of view of one of the bRing cameras during one night for all known passing satellites \emph{(left)} and for known passing satellites in LEO \emph{(right)}, as a function of time. The contour of both figures show the effect of Earth's shadow, as we are able to detect more satellites during evening and morning twilight.}
  \label{fig:observed_satellites}
\end{figure*}

We also tested how many satellites that were known to pass through the field of view of a camera we were able to detect. We cleaned the CSpOC database and removed the (small) debris that would not be visible with our optical camera system\footnote{The faintest satellites we have been able to detect with our current routines had a visual (V-band) magnitude of 7 at the moment of observation, which was derived by comparing its brightness to that of the surrounding stars.}. We define the remaining objects from the CSpoC database as detected if we are able to determine their position within 10 pixels from the position derived from the CSpOC TLE. In this way, we might rule out a large portion of satellites that we did detect but not accurately enough. 

The left panel of Figure \ref{fig:observed_satellites} shows the fraction of all known satellites in the field of view of a bRing camera that we detected during one night. On average, we are able to detect roughly 4\,\% of passing satellites within 10 pixels, although our routines have been developed with a focus on long line segments, i.e. LEO satellites. During twilight we are able to observe up to 8\,\% of all passing satellites. 

In the right panel of Figure \ref{fig:observed_satellites} we look at the subset of passing LEO satellites. For the purpose of this analysis, we have defined a satellite to be in LEO if it has more than 10 orbits per day. From the subset, we are able to detect 25\,\% on average with an accuracy of 10 pixels, and up to 50 to 80\,\% during twilight.

\section{Outlook}

In a follow-up to the pilot study presented here, we will pursue the following goals:
\begin{enumerate}
\item Refine our methods
\item Use data from more sensitive instruments
\item Develop characterisation techniques.
\end{enumerate}

We want to obtain a better sensitivity and more precise orbit determination with our current observational set-up. We will refine our routines and extend them to fainter objects and satellites in higher orbits, such as cubesats in LEO and satellites in geosynchronous orbit. We will have access to data of the BlackGEM array of survey telescopes, that uses six optical Sloan filters and is able to go down to 23rd magnitude in 5 minutes \citep{bloemen16}, which is ideal for the observation of geosynchronous satellites. Apart from detecting, we also want to be able to identify and/or characterise overflying objects using spectropolarimetry. Characterization is highly useful in two ways: it provides intrinsic information about the object and it allows better matching of repeated observations of unknown objects, thereby facilitating the determination of their orbits.

We plan to develop a dedicated observational set-up that simultaneously collects the photometric, spectral and polarimetric profile of satellites and debris. Spectropolarimetry as a function of the illumination angle can differentiate in material composition between satellites, as well as identify the differences in structure or texture for a single material. Satellites in geosynchronous orbits are ideal targets. There are many that are large and reasonably bright, they move slowly on sky and can be observed at a wide range of precisely known phase angles. The spectropolarimetric observation of these targets can be done using a modest telescope, e.g. 30 cm in diameter. Using geosynchronous satellites as a test population has the further advantage that most of them are well known, and information about their size, layout and materials composition is available. This potentially makes it possible to make a list of known features and characteristics that are identifiable in the spectropolarimetric profile of a target.

The technique of spectropolarimetry is also highly relevant for the characterization of NEOs. A known challenge for NEO observations is that it is very difficult to determine the size, and thus the damage potential, based on the intensity measurements: an observed object of a given brightness may either be small and highly reflective or large and dark. Experiments \citep{bagnulo15} have shown that spectropolarimetry may be able to resolve this size-albedo ambiguity. The experiments indicated that it was possible to categorize observed objects into several classes of known origin and composition. This in turn allowed for a much more reliable estimate of the albedo and thus the size of the object.

\section{Conclusion}

We have demonstrated that with a relatively simple optical instrument, a large fraction, up to 80\,\%,  of overflying satellites in low-earth orbit can be detected with an accuracy of a few pixels. We have developed routines that can automatically detect the satellite tracks and perform an orbit refinement or initial orbit determination from the gathered data, even if the orbital elements are not known a priori. Satellites with unknown orbital elements require multiple passages before their next overflight can be predicted accurately.

The developed algorithms are able to match the detected tracks with known objects in databases, e.g. CSpOC, and to flag objects not found in the database. The algorithms are generic and can be used on any optical observation. They can be further optimised to enhance the observing capability of satellites in low Earth orbit and extend it to satellites in geosynchronous orbit. 

\section*{Acknowledgements}

This feasibility study was commissioned by the Space Security Center of the Royal Netherlands Air Force.

\bibliographystyle{plain} 
\bibliography{SSC}

\end{document}